\documentclass[a4paper, 10pt]{article}
\usepackage{cite,color}

\newcommand{\ds}[1]{\textcolor{black}{#1}\ }
\newtheorem{theorem}{Theorem}
\newtheorem{postulate}{Postulate}

\begin{document}
\title{The Application of Non-Cooperative Stackelberg Game Theory in Behavioral Sciences: Social Optimality with any Number of Players}


\author{Jie Dong, Nicole Sawyer, and David B. Smith \\
The Australian National University, \\
National ICT Australia (NICTA)\thanks{NICTA is funded by the Australian Government through the Department of Communications and the Australian Research Council through the ICT Centre of Excellence Program.}, ACT, Australia. \\
Email: \{jie.dong, nicole.sawyer, david.smith\}@nicta.com.au}

\date{}

\maketitle

\begin{abstract}
Here \ds{we present a} \ds{ground-breaking} new postulate for game theory. The first part of this postulate \ds{contains the axiomatic observation that all games are created by a designer, whether they are: e.g., (dynamic/static) or (stationary/non-stationary) or (sequential/one-shot) non-cooperative games, and importantly, whether or not they are intended to represent a non-cooperative Stackelberg game, they} can be mapped to a Stackelberg game. \ds{I.e., the game designer is the leader} who is totally rational and honest, and the followers are mapped to the players of the designed game.  \ds{If now the game designer, or ``the leader" in the Stackelberg context, adopts a pure strategy, we postulate the following second part following from axiomatic observation of ultimate game leadership, where empirical insight leads to the second part of this postulate.} \emph{Importantly,} implementing a non-cooperative Stackelberg game, with a very honest and rational leader \emph{results in social optimality} for all players (followers), assuming pure strategy across all followers and leader, and that the leader is totally rational, honest, and is able to achieve a \ds{minimum} amount of competency in leading this game, with any finite number of iterations of leading this finite game.
\end{abstract} \

John Nash stated the following theorem\ds{ for non-cooperative games that became widely known as the Nash equilibrium:,}
\begin{theorem} [Nash Equilibrium]
Every game with a finite number of players and action profiles has at least one Nash equilibrium \cite{nash1951non}.  A mixed\footnote{Mixed implies non-deterministic} strategy always has an equilibrium.\footnote{i.e., a best response to all others' best responses, which is not guaranteed for a completely deterministic, or pure, strategy.} \cite{nash1951non}.
\label{NashEquilibrium}
\end{theorem} \

We have \ds{discovered} a postulate as a counter-point to the Nash-Equilibrium Theorem \ref{NashEquilibrium} above. Our postulate is:
\begin{postulate} [Postulate on Social Optimality for Non-Cooperative Games]
\ds{In any game that has any number of players, $\mathcal{N}, 2 \leq N \leq \infty$, where each player has a finite deterministic action profile (i.e., pure strategy), then social optimality is potentially achievable, requiring only the following two conditions: All players are being led by a totally rational and honest leader, resulting} in a unique Stackelberg, \cite{von2010market}, equilibrium. \ds{Then the rational, honest leader may be able to implement a dynamic game to achieve a fully hierarchical equilibrium, where the action profiles from both leader and followers are deterministic. Such that assuming competency, honesty and rationality --- social optimality is achievable across all players with a competent and rational leader, or equivalently, a suitably intelligent, game designer}.\footnote{This is even if there are any arbitrary number of players demonstrated varying degrees of irrationality/rationality.}
\label{Postulate1}
\end{postulate} \

Expanding on Postulate \ref{Postulate1}, there are a finite to an infinite number of players which have pure strategies.  Due to the players having pure strategies, the Nash equilibrium is not guaranteed, as stated in Theorem \ref{NashEquilibrium}.  But, however, we have an empirical observation from a test-case of ``global" Stackelberg game design in transmit radio power control for multiple wireless body area network (BANs) coexistence, where a hierarchical equilibrium is ensured by the followers and this was then iterated by the principal game designer to achieve a socially optimal outcome for the BAN coexistence, despite that all previous theory suggested that this was not possible for any wireless communications network with power control based on a non-cooperative game \cite{Dong_ACM2015}. I.e., in a behavioral science context of the game designer, principally J. Dong, providing competent and intelligent leadership for the followers game of distributed transmit power control with no central coordination, then social optimality was postulated empirically from results of implementation.

To obtain Postulate \ref{Postulate1}, a series of papers were used as tools to solve this property.  The first paper which sparked this postulate was a non-cooperative transmit power control game\cite{smith2014multi}, which obtained Pareto-efficient outcomes such as minimized transmit power and rapid convergence to target packet delivery ratios (PDRs).  Next, \cite{smith2014multi} was simplified for a standard proposal to the IEEE 802.15 task group 8 \cite{standard}, which is currently being considered \ds{for the draft 802.15.8 specification}.  The utility function from \cite{smith2014multi} was modified such that it had strict concavity in \cite{Dong_ACM2015} and under all conditions guaranteed a unique Nash equilibrium.  In \cite{tushar2012economics}, a Stackelberg game was proposed, \ds{showing that infinite number of players can potentially be mapped into four-playing group types}. Finally it can be inferred from \cite{wright2014level}, \ds{crucially to behavioral sciences,} \ds{that followers could potentially mapped into four levels of rationality in games}, which \ds{if maintained with properly initiated pure strategy action profiles from the game leader,} then \emph{We postulate, that in the general case, social optimality is still achievable.}

This is considering a limit of four possible categories of game-theoretic behavior: (0) Totally Irrational \cite{wright2014level}; (1) More irrational than rational (i.e., somewhat irrational); (2) More rational than irrational (i.e., Somewhat rational) and (3) Totally rational; that encapsulates all possible game behaviors for non-rational games.

\bibliography{references}

\begin{thebibliography}{}

\bibitem[Dong et~al., 2015]{Dong_ACM2015}
Dong, J., Smith, D., and Hanlen, L. (2015).
\newblock Socially optimal coexistence of wireless body area networks enabled
  by a non-cooperative game.
\newblock arXiv:submit/1242776[cs:N1].

\bibitem[Nash, 1951]{nash1951non}
Nash, J. (1951).
\newblock Non-cooperative games.
\newblock {\em Annals of mathematics}, pages 286--295.

\bibitem[Smith et~al., 2014a]{smith2014multi}
Smith, D.~B., Portmann, M., Tan, W.~L., and Tushar, W. (2014a).
\newblock Multi-source--destination distributed wireless networks:
  Pareto-efficient dynamic power control game with rapid convergence.
\newblock {\em Vehicular Technology, IEEE Transactions on}, 63(6):2744--2754.

\bibitem[Smith et~al., 2014b]{standard}
Smith, D.~B., Sawyer, N., Zhou, S., Hernandez, M., Li, H.-B., Dotlić, I., and
  Miura, R. (2014b).
\newblock Proposal outline of completely distributed power control mechanism
  for peer-aware communications.
\newblock IEEE Standard P802.15 Working Group for Wireless Personal Area
  Networks (WPANs).

\bibitem[Tushar et~al., 2012]{tushar2012economics}
Tushar, W., Saad, W., Poor, H.~V., and Smith, D.~B. (2012).
\newblock Economics of electric vehicle charging: A game theoretic approach.
\newblock {\em Smart Grid, IEEE Transactions on}, 3(4):1767--1778.

\bibitem[Von~Stackelberg, 2010]{von2010market}
Von~Stackelberg, H. (2010).
\newblock {\em Market structure and equilibrium}.
\newblock Springer Science \& Business Media.

\bibitem[Wright and Leyton-Brown, 2014]{wright2014level}
Wright, J.~R. and Leyton-Brown, K. (2014).
\newblock Level-0 meta-models for predicting human behavior in games.
\newblock In {\em Proceedings of the fifteenth ACM conference on Economics and
  computation}, pages 857--874. ACM.

\end{thebibliography}
\bibliographystyle{apalike}

\end{document}